\documentclass[ 
    aps,                
    prd,                
    onecolumn,         
    12pts,              
    groupedaddress,     
    superscriptaddress, 
    preprintnumbers,    
    floatfix,
    nofootinbib,        
    notitlepage,          
    showkeys,           
    showpacs,           
    ]{revtex4-2}

\usepackage[utf8]{inputenc}
\usepackage{epsfig,amsfonts,amsthm}
\usepackage{epstopdf}
\usepackage{amsfonts,amsmath,amssymb}
\usepackage{gensymb}
\usepackage{latexsym}
\usepackage{bbm}
\usepackage{array}
\usepackage{comment}
\usepackage{hyperref}
\usepackage{color}
\usepackage{pb-diagram}
\usepackage{slashed}
\usepackage{multirow}
\usepackage{cancel}
\usepackage{placeins}                       
\usepackage{diagbox}

\newcommand{\be}{\begin{equation}}
\newcommand{\ee}{\end{equation}}
\newcommand{\bea}{\begin{eqnarray}}
\newcommand{\eea}{\end{eqnarray}}
\newcommand{\diag}{{\rm{diag}}}

\newcommand{\mean}[1]{\left\langle#1\right\rangle}
\newcommand{\unitv}[1]{\hat{#1}}
\newcommand{\difd}{\,d}

\def\lsim{\mathrel{\raise.3ex\hbox{$<$\kern-.75em\lower1ex\hbox{$\sim$}}}}
\def\gsim{\mathrel{\raise.3ex\hbox{$>$\kern-.75em\lower1ex\hbox{$\sim$}}}}

\begin{document}

\title{Probing the dark matter velocity distribution via daily modulation}

\author{Matti Heikinheimo}
\email{matti.heikinheimo@helsinki.fi}
\affiliation{Department of Physics, University of Helsinki, 
                      P.O.Box 64, FI-00014 University of Helsinki, Finland}
\affiliation{Helsinki Institute of Physics, 
                      P.O.Box 64, FI-00014 University of Helsinki, Finland}

\author{Kai Nordlund}
\email{kai.nordlund@helsinki.fi}
\affiliation{Department of Physics, University of Helsinki, 
                      P.O.Box 64, FI-00014 University of Helsinki, Finland}
\affiliation{Helsinki Institute of Physics, 
                      P.O.Box 64, FI-00014 University of Helsinki, Finland}

\author{Sebastian Sassi}
\email{sebastian.k.sassi@helsinki.fi}
\affiliation{Department of Physics, University of Helsinki, 
                      P.O.Box 64, FI-00014 University of Helsinki, Finland}
\affiliation{Helsinki Institute of Physics, 
                      P.O.Box 64, FI-00014 University of Helsinki, Finland}

\author{Kimmo Tuominen}
\email{kimmo.i.tuominen@helsinki.fi}
\affiliation{Department of Physics, University of Helsinki, 
                      P.O.Box 64, FI-00014 University of Helsinki, Finland}
\affiliation{Helsinki Institute of Physics, 
                      P.O.Box 64, FI-00014 University of Helsinki, Finland}

\begin{abstract}
\noindent
{We consider dark matter velocity distributions with an anisotropic component, and analyze how the velocity structure can be probed in a solid state ionization detector with no directional detection capability using a daily modulation effect due to the anisotropic response function of 
the target. We show that with an energy resolution of $<10$ eV it is possible to identify the presence of an anisotropic component consistent with observations for sub-GeV dark matter, and that introduction of daily modulation information substantially improves the sensitivity in a narrow mass range.}
 \end{abstract}
\preprint{HIP-2023-17/TH}
\maketitle

\section{Introduction}
\label{sec:intro}

The paradigm of cold dark matter (DM) explains, within the Standard Model of cosmology, the observed features of cosmic microwave background, large scale structure in the universe and its hierarchical formation~\cite{Planck:2018vyg, Reid:2009xm,Tinker:2011pv}. While astrophysical and cosmological observations unravel the abundance of dark matter, they do not allow insight into more detailed properties of a hypothetical dark matter particle.  

Even the most basic parameter of dark matter particle, its mass, remains unknown. The low end is set around $m\sim 10^{-22}$ eV by ultralight bosons~\cite{Turner:1983he,Press:1989id,Sin:1992bg} having de Broglie wavelength of $\lambda\sim 1$ kpc. Together with the high end around ${\cal O}(10\,M_\odot)$, this leads to a dynamical mass range of hundred orders of magnitude. The truly vast mass range is reflected in a great variety of different dark matter candidates and their interaction strengths.

It is an appealing possibility that DM would be constituted by a new elementary particle, embedded with ordinary matter in some theory beyond the Standard Model (SM) of elementary particle interactions. This allows to determine the present DM abundance by processes in the early universe, such as an asymmetry~\cite{Hut:1979xw,Nussinov:1985xr,Barr:1990ca}, thermal decoupling~\cite{Kolb:1990vq,Gondolo:1990dk} or freezing in~\cite{McDonald:2001vt,Bernal:2017kxu}.  For cold particle DM whose origin is in thermal processes, the DM mass range is constrained from below, $m\gsim {\cal O}(\text{keV})$, by structure formation and from above, $m\lsim{\cal O}(100 \text{ TeV})$, by unitarity of the scattering cross section. 

One simple possiblity of keV-scale DM is sterile neutrinos~\cite{Drewes:2016upu}.
As another example, many beyond SM scenarios give rise to Weakly Interacting Massive Particles (WIMPs), which have mass in the range $m\gsim 10$ GeV to few TeV and couple to the Standard Model via the electroweak interaction~\cite{Steigman:1984ac}. Recently, in a more general setting, much theoretical and phenomenological interest has focused on various hidden sector models where the particle content is singlet under the Standard Model interactions and couples only through specific portal interactions~\cite{Silveira:1985rk,McDonald:1993ex,Fradette:2014sza,Bai:2014osa}. Often the DM candidates in this context are also referred to as 'WIMP', where the weakness means the overall strength of the non-gravitational interaction. Such general models provide simple benchmarks over the mass range ${\cal O}(\text{keV}-\text{TeV})$ to be searched for in direct detection experiments.   

The methods to directly detect interactions between dark matter and ordinary matter depend on the DM mass range and its interactions. For cold particle matter with masses above keV-scale,
the direct detection of dark matter 
attempts to 
observe the small recoil energy deposits in scattering events between DM particles and some target material. 
Experiments utilizing liquid xenon target have made great progress over the past decades in constraining dark matter interactions as a function of dark matter mass~\cite{aprile2017first,cui2017dark, wang2020results,meng2021dark,lux2023first,aprile2023first}. Scattering on Xenon atoms rapidly loses sensitivity below dark matter masses of 10 GeV. To constrain dark matter with sub-GeV mass one can try to utilize sensitivity to directional dependence of the dark matter scattering evenets~\cite{Battat:2016pap}. This can be expected to be applicable in particular in detectors using crystals as a target, e.g.~\cite{SuperCDMS:2017nns,SuperCDMS:2018mne,SuperCDMS:2020ymb}.

Generally, the expected rate for scattering events with given observable characteristics depends on the scattering cross section between the DM and the target atoms, on the flux of the DM particles incident on the target, and on the response function of the target material~\cite{Trickle:2019nya}.
The latter is a property specific to the material, describing the many body effects relevant for transforming the initial nuclear or electron recoil into an observable signal with given measurable properties. The DM flux is determined by the velocity distribution of the DM particles in the rest frame of the detector, i.e. the laboratory frame. 

For light dark matter with mass in the range of $\lsim 1$ GeV, the response function of solid state target materials plays an important role in understanding the signal event rate, as discussed recently e.g. in \cite{Trickle:2019nya, Trickle:2020oki, Mitridate:2022tnv, Essig:2022dfa, Boyd:2022tcn}. An example of such phenomena is the anisotropic threshold for nuclear recoil induced ionization in a semiconductor target, discussed in \cite{Kadribasic:2017obi, Heikinheimo:2019lwg}. This effect leads to a daily modulation in the event rate, and also affects the annual modulation of the event rate \cite{Sassi:2021umf}.

The dark matter velocity distribution is commonly assumed to be an equilibrium distribution in the galactic rest frame, where the velocity dispersion is given by the virial temperature of the Milky Way (MW) halo, with a high velocity cut-off removing particles with velocities above the local escape velocity. This distribution, called the standard halo model (SHM), would presumably be the result of a perfect virialization. However, from observations of stars in the Milky Way (MW), it is evident that the MW is not perfectly virialized, and therefore it is reasonable to assume that neither is the DM halo. Based on the GAIA measurements \cite{Gaia:2018ydn} of the motions of MW stars, Evans et al. \cite{Evans:2018bqy} presented an updated halo model SHM$^{++}$ that features an anisotropic 'sausage' component and discussed how this updated halo model would alter the predictions for DM event rate in direct detection experiments. Other non-SHM velocity distributions and their effect in dark matter direct detection have been discussed in the literature, see e.g.~\cite{Lee:2012pf,Kavanagh:2016xfi,Mandal:2018efq,Necib:2018iwb,OHare:2018trr,Besla:2019xbx,OHare:2019qxc,Buch:2019aiw,Bozorgnia:2019mjk,Smith-Orlik:2023kyl,Arza:2022dng,Kryemadhi:2022vuk,Maity_2023}.

In this paper we analyze the expected daily and annual modulation signals in a low threshold germanium ionization detector in the presence of an anisotropic component in the velocity distribution of the MW halo. We model the response of the detector by approximating the directional dependence of ionization by the threshold energy for defect creation~\cite{Kadribasic:2017obi, Heikinheimo:2019lwg}. This is determined by first principle molecular dynamics, see e.g.~\cite{PhysRevB.78.045202,NORDLUND2006322} for a detailed discussion and~\cite{Kadribasic:2017obi, Heikinheimo:2019lwg,Sassi:2021umf} for earlier applications in the context of dark matter detection. We show that at high degrees of anisotropy, realized in the SHM$^{++}$ model, the angular distribution of dark matter scattering events from the anisotropic component exhibits a bimodal structure, which amplifies the high frequency components of the daily modulation signal relative to that produced by a fully isotropic distribution. We also perform a likelihood ratio test analysis to study the strength of direct detection signal needed to detect the presence of an anisotropic component in the velocity distribution.

We find in particular, that with a sufficiently large number of signal events, and for DM masses for which most of the recoil spectrum is above the detector threshold, an experiment with $<10$ eV energy resolution is sensitive to the features of the SHM$^{++}$ model. As the DM masses approach the values at the detection threshold, the gains in sensitivity to the  anisotropic component, although substantial, are not sufficient to probe the SHM$^{++}$ region.

The paper is organized as follows: In section~\ref{sec:rate} we review the basic formulas for the event rate, emphasing the structure of integrals over velocity in the case of directional sensitivity of the target. In section~\ref{sec:angint} we describe the numerical methods we have applied to perform the angular integrals in the event rate and in sections
\ref{sec:modulation} and \ref{sec:sensitivity} we discuss the periodic structure of the event rate and how this affects the sensitivity of a semiconductor detector to the anisotropic component in the SHM$^{++}$ DM velocity distribution. In section~\ref{sec:checkout} we present our conclusions.   

\section{Dark matter event rate and velocity distributions}
\label{sec:rate}

Let us start by reviewing briefly the main formulas relevant for scattering of a DM particle on a target nucleus. The differential DM--nucleus scattering rate (per unit detector mass) is given by~\cite{Gondolo:2002np}
\begin{equation}
\frac{d^2R_S}{dEd\Omega}=\frac{1}{64\pi^2}\frac{\rho_0}{m_\text{DM}^3m_\text{N}^2}\int\mean{|\mathcal{M}|^2}\delta(\vec{v}\cdot\unitv{q}-v_\text{min})f(\vec{v})\difd^3 v.
\label{eq:dmrate}
\end{equation}
Here $\rho_0$ is the local dark matter density, $m_\text{DM}$ and $m_\text{N}$ are the masses of the dark matter particle and nucleus, respectively, $\mathcal{M}$ is the DM--nucleus scattering amplitude, and $f(\vec{v})$ is the dark matter velocity distribution. The speed $v_\text{min}$ is defined as 
\begin{equation}
v_\text{min}=\sqrt{\frac{m_\text{N}E}{2\mu_\text{DM,N}^2}},
\end{equation}
where $E$ is the energy of the scattering nucleus and $\mu_\text{DM,N}$ is the reduced mass of the DM--nucleus system. Commonly the interaction is assumed to be straightforward spin-independent scattering, in which case the scattering amplitude may be simply replaced with a cross section. In this work we describe the scattering amplitude in the more general framework of a non-relativistic effective theory \cite{Fitzpatrick:2012ix} where all interactions are expressible as combinations of twelve operators. From this theory follows the result that
\begin{equation}
\mean{|\mathcal{M}|^2}=\frac{m_\text{N}^2}{m_n^2}\sum_{j,k=1}^{12}\sum_{a,b=0}^1c_j^ac_k^b W_{jk}^{ab}(q^2,(\vec{v}_\perp)^2),
\end{equation}
where $W_{jk}^{ab}(q^2,(\vec{v}_\perp)^2)$ depend on the nuclear form factors, and are functions of the momentum transfer $q^2$
and the DM velocity perpendicular to the momentum transfer, $(\vec{v}_\perp)^2$. The factors $c_j^a$ are coefficients of the interaction operators in the effective theory with $a=0,1$ an isospin index (corresponding to isoscalar and isovector). In the limit of small $q$, the form factors can be taken as truncated polynomials. In fact, usually the lowest order term is sufficient. Then $W_{jk}^{ab}(q^2,(\vec{v}_\perp)^2)$ reduce to polynomials of $q^2$ and $(\vec{v}_\perp)^2$. It follows that we can transform the sum into a form
\begin{equation}
\mean{|\mathcal{M}|^2}=\sum_{n=0}^\infty\sum_{s=0}^1M_{ns}q^{2n}(\vec{v}_\perp)^{2s}.
\label{eq:amplitudepoly}
\end{equation}
The main consequence of these considerations is that the computation of the integral \eqref{eq:dmrate} reduces to the computation of two Radon transform integrals
\begin{equation}
\hat{f}_s(\unitv{q},v_\text{min})=\int(\vec{v}_\perp)^{2s}\delta(\vec{v}\cdot\unitv{q}-v_\text{min})f(\vec{v})\difd^3 v,
\label{eq:radon}
\end{equation}
with $s=0$ or $s=1$.

For the dark matter velocity distribution, the standard halo model (SHM) is usually assumed, which has a simple isotropic form
\begin{equation}
f_\text{I}(\vec{v})=\frac{1}{N}e^{-v^2/v_0^2}\Theta(v_\text{esc}-v).
\label{eq:isodist}
\end{equation}
Here $N$ is a normalization constant, $v_0$ is the velocity dispersion, which is taken equal to the local circular speed of the Milky Way at the location of the solar system, and the escape velocity of the Milky Way $v_\text{esc}$ acts as an upper limit for dark matter velocities.

This simple isotropic approximation is easy to work with, but unlikely to be the correct shape of the velocity distribution. Recent studies suggest at least the presence of an anisotropic component in addition to the isotropic one, which gives rise to the SHM$^{++}$ model presented in \cite{Evans:2018bqy}. The form of this 
velocity distribution is
\be
f_{{\rm SHM}^{++}}(\vec v) = (1-\eta)f_\text{I}(\vec v) + \eta f_\text{A}(\vec v),
\label{eq:SHM++}
\ee
where $f_\text{I}$ is the isotropic component \eqref{eq:isodist}, $f_\text{A}$ is the anisotropic component and $0 \leq \eta \leq 1$ is the anisotropy fraction. The velocity distribution for the anisotropic component is
\be
f_\text{A}(\vec v)=\frac{1}{N}e^{-\frac{1}{2}\vec{v}^T\Sigma^{-1}\vec{v}}\Theta(v_{\rm{esc}}-v),
\label{eq:anisodist}
\ee
where $\Sigma=\diag(\sigma_r^2,\sigma_\theta^2,\sigma_\phi^2)$ in the galactic frame, with
\be
\sigma_r^2=\frac{3v_0^2}{2(3-2\beta)},\quad\sigma_\theta^2=\sigma_\phi^2=\frac{3v_0^2(1-\beta)}{2(3-2\beta)}.
\label{eq:dispersion}
\ee
From~\cite{Evans:2018bqy}, the value of the circular rotation speed in SHM$^{++}$ is $v_0=233$ km s$^{-1}$, the escape velocity is $v_{\rm{esc}}=528$ km s$^{-1}$ and the anistropy parameter  $\beta$ is 0.9.

The scattering of dark matter particles is not observed in the rest frame of the velocity distribution, which is the galactic frame, but usually in an Earth-based laboratory. Therefore in equations \eqref{eq:dmrate} and \eqref{eq:radon} the velocity $\vec{v}$ is replaced with $\vec{v}+\vec{v}_\text{lab}$, where $\vec{v}_\text{lab}$ is the velocity of the laboratory in the galactic frame. It can be expressed in terms of its various contributions as $\vec{v}_\text{lab}=\vec{v}_\text{circ}+\vec{v}_\text{pec}+\vec{v}_\text{earth}+\vec{v}_\text{rot}$ \cite{Mayet:2016zxu}. Here $\vec{v}_\text{circ}$ is the circular velocity of the solar system in the galactic plane, $\vec{v}_\text{pec}$ is the peculiar velocity, $\vec{v}_\text{earth}$ is the velocity of the Earth around the sun, and $\vec{v}_\text{rot}$ is the rotational velocity of the point on Earth's surface. The variation in the magnitude of $\vec{v}_\text{lab}$ due to the changing direction of $\vec{v}_\text{earth}$ results in the well known annual modulation of the dark matter event rate.

The SHM distribution is simple enough that the integrals in equation \eqref{eq:radon} are straightforward to evaluate analytically. However, even for the slightly more general anisotropic distribution  \eqref{eq:anisodist}, no analytical expression for the Radon transform exists, and one must resort to numerical evaluation of the integrals. The general numerical integration of dark matter event rates is challenging, because obtaining the complete angle and energy integrated scattering rate requires computation of a six-dimensional integral: first a three-dimensional integral over the velocities of the dark matter particles, and then a two-dimensional integral over the recoil direction of the scattered nucleus and a one-dimensional integral over its energy.

To address the above issues, we present here a simple framework that we have used for efficient evaluation of dark matter event rates. The first observation is the one made in relation to 
Eq.~\eqref{eq:amplitudepoly}, where the effective theory of dark matter enables writing the double differential rate in Eq.~\eqref{eq:dmrate} to be expressed in terms of the Radon integrals from Eq.~\eqref{eq:radon}. This gets rid of any dependence on the model of dark matter interaction within the velocity integral. In the general case where we are not in the rest frame of the dark matter velocity distribution, the velocity distribution $f(\vec{v}+\vec{v}_\text{lab})$ depends on time via $\vec{v}_\text{lab}$, but this can be resolved: the change of the frame can be performed after the Radon transformation~\cite{Gondolo:2002np, Heikinheimo:2019lwg}. Consider a simple change of variables to write the Radon integrals in the form
\begin{equation}
\hat{f}_s(\unitv{q},v_\text{min})=\int((\vec{v}-\vec{v}_\text{lab})_\perp)^{2s}\delta(\vec{v}\cdot\unitv{q}-w)f(\vec{v})\difd^3 v,
\end{equation}
where $w=v_\text{min}+\vec{v}_\text{lab}\cdot\unitv{q}$. The nontransverse ($s=0$) integral is a function of $w$ and $\unitv{q}$, which form a three-dimensional parameter space. The transverse ($s=1$) integral depends also on the components of $\vec{v}_\text{lab}$, which is not desirable. However, we may expand the problematic term as
\begin{equation}
((\vec{v}-\vec{v}_\text{lab})_\perp)^2=v^2-(\unitv{q}\cdot\vec{v})^2+2((\unitv{q}\cdot\vec{v}_\text{lab})\unitv{q}-\vec{v}_\text{lab})\cdot\vec{v}+v_\text{lab}^2-(\unitv{q}\cdot\vec{v}_\text{lab})^2,
\end{equation}
which allows us to write
\begin{equation}
\hat{f}_1(\unitv{q},w)=\hat{f}_1^0(w,\unitv{q})+2((\unitv{q}\cdot\vec{v}_\text{lab})\unitv{q}-\vec{v}_\text{lab})\cdot\hat{\vec{f}}(w,\unitv{q})+(v_\text{lab}^2-(\unitv{q}\cdot\vec{v}_\text{lab})^2)\hat{f}_0(w,\unitv{q}).
\label{eq:transradonsum}
\end{equation}
Here $\hat{f}_1^0(w,\unitv{q})$ is $\hat{f}_1(\unitv{q},w)$ at $\vec{v}_\text{lab}=0$, and $\hat{\vec{f}}(w,\unitv{q})=\int\vec{v}\delta(w-\unitv{q}\cdot\vec{u})f(\vec{v})\difd^3 v$. Importantly, all the integrals are now independent of $\vec{v}_\text{lab}$, only depending on $w$ and $\unitv{q}$. It is therefore possible to construct a collection of three-dimensional interpolants out of these integrals for $w$ and two angle parameters. Computing the Radon transforms then consists of evaluating the relevant interpolants at the point corresponding to $(w,\unitv{q})$, and evaluating the sum \eqref{eq:transradonsum} for the transverse ($s=1$) case. The integrals only need to be recomputed when changing parameters of the velocity distribution, and the results may be cached for reuse. Because of the $\delta$-function appearing in the Radon transform, the integrals that need to be evaluated numerically are only two-dimensional.

When it comes to event rates, we observed that it is generally useful to obtain the differential rate $dR_S/dE$ by first performing the angular integral over the recoil direction. There are two reasons for this: First, $dR_S/dE$ can be expressed as a linear combination of integrals  of the velocity distributions over scattering angle in the Radon transforms. Therefore these two angular integrals can be evaluated for a range of energies, the results can be saved
and repeatedly used for different models of dark matter interaction. Second, with detectors that can measure energy, we are in any case generally interested in the energy spectrum of the event rate, and if $dR_S/dE$ has been evaluated on an energy grid, it is trivial to integrate it to obtain the total rate.

\section{Event rate angle integrals}
\label{sec:angint}

In general, not all recoil events are detectable in a given experiment, given that the recoil needs to result in a signal measurable by the detector hardware. For example, in a semiconductor ionization detector, a nuclear recoil would have to produce an ionization event in the lattice that would then be measured. We therefore express the observable signal event rate in a very general form as
\begin{equation}
\frac{d^2R}{dEd\Omega}=\frac{d^2R_S}{dEd\Omega}P(E,\unitv{q}),
\label{truerate}
\end{equation}
where $R_S$ is the rate of scattering events as defined in Eq.~\eqref{eq:dmrate}, and $P(E,\unitv{q})$ describes the probability that a scattering at energy $E$ in the direction $\unitv{q}$ produces a detectable signal. In materials such as single crystal semiconductors, the function $P(E,\unitv{q})$ may have significant directional dependence. In the context of a dark matter direct detection on Earth, where the scattering rate $R_S$ itself has some anisotropy, this will lead to a signal that changes depending on the detector orientation, which then results in modulation effects as the detector moves along with the Earth as time passes.

For the purpose of this analysis, we use a model where $P(E,\unitv{q})$ is a simple step function $\Theta(E-E_\text{min}(\unitv{q}))$, where $E_\text{min}(\unitv{q})$ is the minimum nuclear recoil energy needed to produce an ionization event. The calculation of the threshold ionization energy $E_\text{min}(\unitv{q})$ from first principles is not feasible with the existing numerical methodology. We use an approximation that the ionization energy in a given direction should be correlated with the crystal defect creation threshold, and therefore use the defect creation threshold energy surface to model $E_\text{min}(\unitv{q})$ as we have done in previous studies~\cite{Kadribasic:2017obi, Heikinheimo:2019lwg,Sassi:2021umf}. This approach is based on the reasoning that the directional dependence of both the ionization and defect creation follow from the symmetries of the crystal structure. The defect creation threshold energies may be computed using well-studied methods of classical molecular dynamics simulations, described in \cite{Heikinheimo:2022ler}. From these simulations we obtain a set of samples $E_\text{min}(\unitv{q}_i)$ of the defect creation threshold energies for a random collection of directions $\{\unitv{q}_i\}$, sampled uniformly on the unit sphere. In principle, equation \eqref{truerate} may be integrated simply by performing a Monte Carlo integral over the sampled directions $\{\unitv{q}_i\}$. This, however, is not ideal due to two issues. 

The first issue is that the final result may be affected by discretization artifacts.
These can be expected to arise because, due to the numerical cost of molecular dynamics simulations, the data contains relatively few directions ($\mathcal{O}(10^5)$ at most). This is generally not an issue if the angular distribution $d^2R/dEd\Omega$ covers a substantial portion of the unit sphere. However, in the specific case of light dark matter with scattering energies at the threshold 
\be 
E\sim \min_{\unitv{q}\in S^2}\{E_\text{min}(\unitv{q})\}, 
\ee 
the angular distribution of recoil events becomes focused to a small section of the sphere, which may be covered by $\mathcal{O}(100)$ of the sample directions. This, then, may lead to spurious modulation effects in the angle integrated signal as the modulation is not only sensitive to the true shape of the energy surface, but also to the random sampling of points. The second issue is that Monte Carlo integration is inherently slow for a given accuracy relative to more sophisticated methods, and is not generally viable for sampling the event rate for thousands of time points with $\mathcal{O}(10^5)$ directions in the angle integral, in the general case where $d^2R_S/dEd\Omega$ has no simple analytic expression.

In order to have a flexible method for computing $d^2R/dEd\Omega$ and its integrals, we use the sampled values $E_\text{min}(\unitv{q}_i)$ to construct a spherical harmonic fit as an approximation of the threshold energy surface. This has the benefit of suppressing spurious high frequency noise. For purposes of numerical efficiency, an interpolating function of the resulting approximation is then constructed. The construction of an approximation of the surface that can be evaluated at any point enables the application of generally more efficient adaptive numerical integration methods on \eqref{truerate}. However, because the support of $\Theta(E-E_\text{min}(\unitv{q}))$ essentially defines the integration region as some complicated subset of the unit sphere, naive integration over the entire sphere may fail due to discontinuities. This is especially true in cases where the support is very small. In such a case all samples of the integrand, selected by the adaptive algorithm in some subregion, might fall on points where the integrand is zero, yielding an erroneous ``exact'' estimate of zero for the integral in the subregion. To mitigate issues with the discontinuities, before evaluating the angular integral, we construct a set of rectangular integration subregions to approximate the support of $\Theta(E-E_\text{min}(\unitv{q}))$, such that the integration routine is guaranteed to properly sample every patch where the integrand is nonzero. The choice of rectangular subdivision here is suboptimal because of the sharp dropoff of the integrand at the boundary, and significant speedups could probably be achieved by instead implementing an integration routine based on triangular subdivision.

As can be seen from observing the molecular simulation data for germanium, the defect creation threshold surface generally contains some features at fairly small angular scales, suggesting that a fully accurate fit would require a relatively high order $\ell_\text{max}$ of spherical harmonic expansion. However, because the angular distribution of dark matter scattering events is relatively more spread out, we find that beyond about $\ell_\text{max}=30$ introducing higher order spherical harmonics does not significantly change the results obtained from the angular integrals.

\begin{figure}[h]
\includegraphics{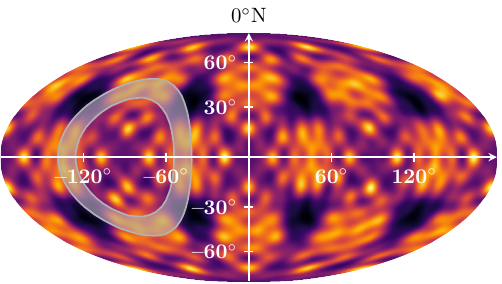}
\includegraphics{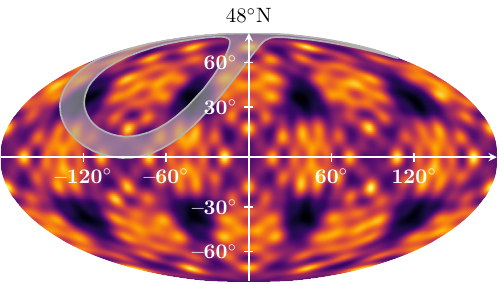}
\caption{Directions swept by $-\vec{v}_\text{lab}$ throughout the year (shaded band) at latitudes 0\degree{} and 48\degree{} north overlaid on the energy threshold surface of a germanium crystal. The angles are given in horizontal coordinates, i.e. (0\degree, 90\degree) is towards zenith, and (0\degree, 0\degree) is south. The crystal is oriented such that the faces of the rectangular unit cell are perpendicular to the coordinate axes.}
\label{fig:dm_wind_dir}
\end{figure}

\section{Structure of event rate modulation}
\label{sec:modulation}

\begin{figure}[h]
\includegraphics{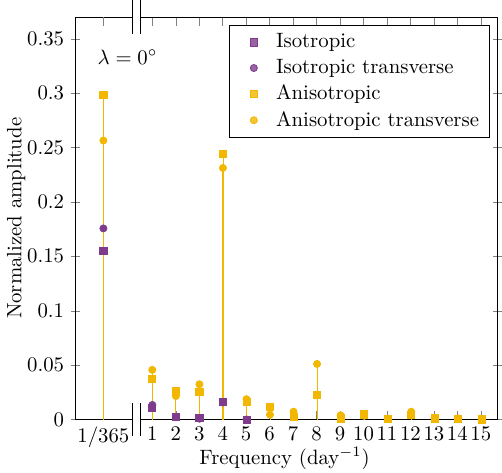}
\includegraphics{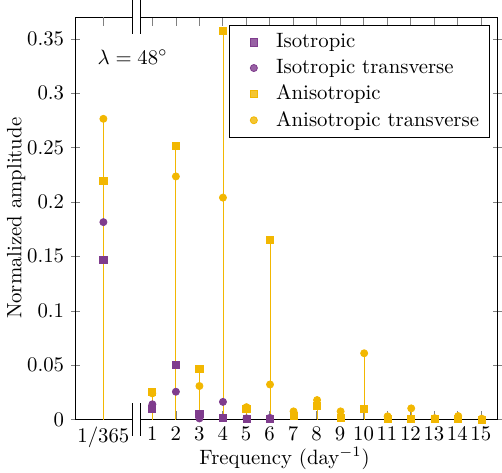}
\caption{Fourier spectra for the isotropic and anisotropic contributions to the event rate at 0\degree and 48\degree, respectively, both for the transverse and nontransverse spin-independent EFT operators. The normalization is relative to the zero-frequency component, i.e. the average rate.}
\label{fig:fourier_spectrum}
\end{figure}

The anisotropy of the detector material is what gives rise to a daily modulation effect. We therefore expect the precise nature of the modulation signal to depend on the structure of the anisotropy. In particular, in crystalline detector materials, where the anisotropy results from the crystal structure, the structure of the modulation signal is sensitive to the symmetries of the crystal structure as well as to the orientation of the crystal in the laboratory frame. It is additionally dependent on the dark matter velocity distribution itself, as well as the velocity $\vec{v}_\text{lab}$ of the laboratory frame relative to the velocity distribution, which determine the angular distribution of the dark matter scattering rate in the laboratory frame.

Figure \ref{fig:dm_wind_dir} shows the region of the sky swept by the dark matter wind direction ($-\unitv{v}_\text{lab}$) throughout the year in two different coordinate systems, corresponding to two different latitudes: at the equator (0\degree{} north), and at 48\degree{} north. Each day the rotation of the Earth causes the direction of the dark matter wind to sweep a circular path around the north pole of the celestial sphere, and the radius of the path changes gradually throughout the year due to the motion of the Earth around the Sun. In the figure, the directions have been overlaid on top of the defect creation threshold energy surface of germanium. The darker areas of the surface correspond to minima of the threshold energy surface, and therefore to directions in which more scattering events are detected. We can see that if the scattering rate of dark matter was very localized around the direction of the dark matter wind, e.g., if the velocity distribution was very stream-like, then we could expect the number of minima in the event rate signal during a single day to correspond to the number of energy surface minima crossed by the dark matter wind direction throughout the day. That is, in the equatorial case we would expect the main daily modulation component to have four periods per day, whereas in the 48\degree{} case we would expect it to have two periods per day.

To confirm this, we performed a Fourier analysis of the event rate signal for dark matter of mass 340 MeV for a detector at different latitudes. This analysis was performed for the isotropic and anisotropic components of the SHM$^{++}$ velocity distributions separately (i.e. cases $\eta=0$ and $\eta=1$ in equation \eqref{eq:SHM++}) to showcase the differences in the modulation effect between different velocity distributions.

\begin{figure}[h]
\includegraphics{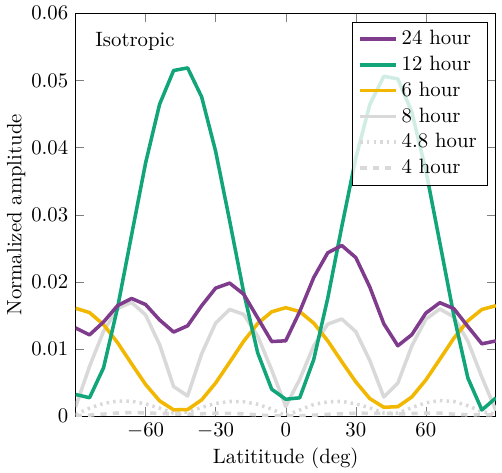}
\includegraphics{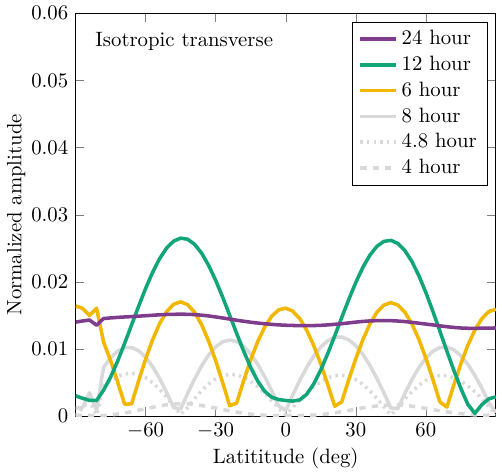}
\includegraphics{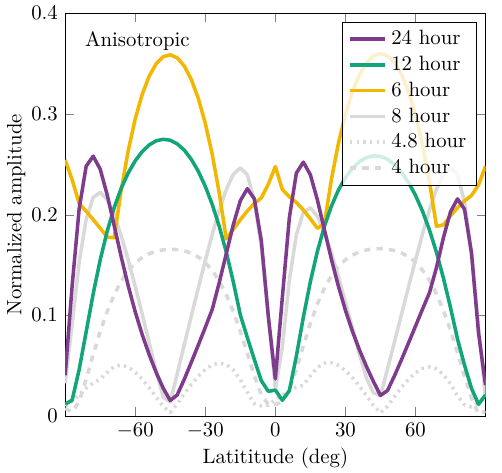}
\includegraphics{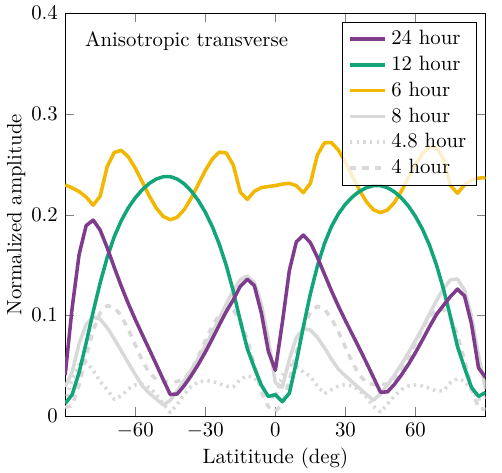}
\caption{Amplitudes of Fourier components plotted as a function of latitude for the isotropic and anisotropic contributions to the event rate, both for the transverse and nontransverse spin-independent EFT operators. The normalization is relative to the zero-frequency component, i.e. the average rate.}
\label{fig:fourier}
\end{figure}

In figure~\ref{fig:fourier_spectrum} we show the Fourier spectrum for two example latitudes from figure~\ref{fig:dm_wind_dir}. The figure shows the spectrum for both the isotropic and anisotropic velocity distribution, and for the transverse and nontransverse Radon transforms. The spectrum shows how different Fourier modes contribute as the orientation of the crystal changes. One can also observe that the amplitude of the anisotropic component is larger than the amplitude of the isotropic one. The dominant Fourier components  are shown in figure~\ref{fig:fourier} as a function of the latitude. In general, the dominant component almost always appears to be one of the components with either a 24 hour, 12 hour, or 6 hour period, with the component with 8 hour period also reaching significant amplitudes. This is consistent with the observations that can be made from figure \ref{fig:dm_wind_dir}, where these periods correspond to touching one, two, four, and three minima, respectively. Specifically, we observe that at latitude 0\degree N, the 12 hour component vanishes almost completely, and reaches its maximum at around 45\degree N, exactly as one would expect. Based on the reflection symmetry of the crystal lattice we would expect a corresponding symmetry in the figure for reflections about zero latitude. However, we notice some deviations from this symmetry, which we suspect originate from the discrete sampling of the time series data. 

A significant difference between the isotropic and anisotropic distributions can be seen in the behavior of the 6 hour component in the nontransverse case, where the component obtains a minimum at 45\degree N for the isotropic distribution, but it obtains a maximum for the anisotropic distribution. This is a unique consequence of highly anisotropic velocity dispersion in the SHM$^{++}$ model. Namely, for values of $\beta$ close to unity in equations \eqref{eq:dispersion}, the dispersion components $\sigma_\theta$ and $\sigma_\phi$ approach zero and $\sigma_r$ increases. Consequently, the isosurfaces of the velocity dispersion become highly elongated ellipses on the radial axis from the galactic center, which means that velocities of most particles have substantial radial components in the galactic frame. As a result, in the lab frame the angular distribution of dark matter events ends up with two peaks with the direction of $-\vec{v}_\text{lab}$ between them. This doubling of the peaks amplifies the higher frequency Fourier components, since daily modulation maxima are obtained when both peaks pass a minimum of the energy surface.

A similar effect is seen in the transverse case relative to the nontransverse case, but for different reasons. Namely, the appearance of $(\vec{v}_\perp)^2$ in the transverse Radon integral implies that nuclear scatterings in the direction of $\vec{v}$ are suppressed, and the angular distribution of scattering events becomes ring-like with a dip where the nontransverse distribution would have a maximum. This more complex shape of the angular distribution again amplifies some of the higher order Fourier components.

\begin{figure}
\includegraphics{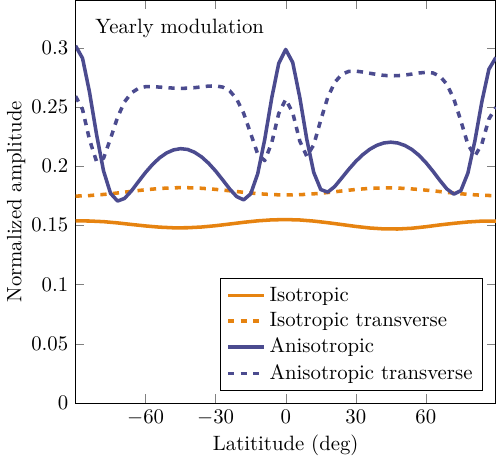}
\caption{Amplitudes of the Fourier components of yearly modulation of the dark matter event rate with the same normalization as in figure \ref{fig:fourier}.}
\label{fig:fourieryearly}
\end{figure}

\begin{figure}
\includegraphics{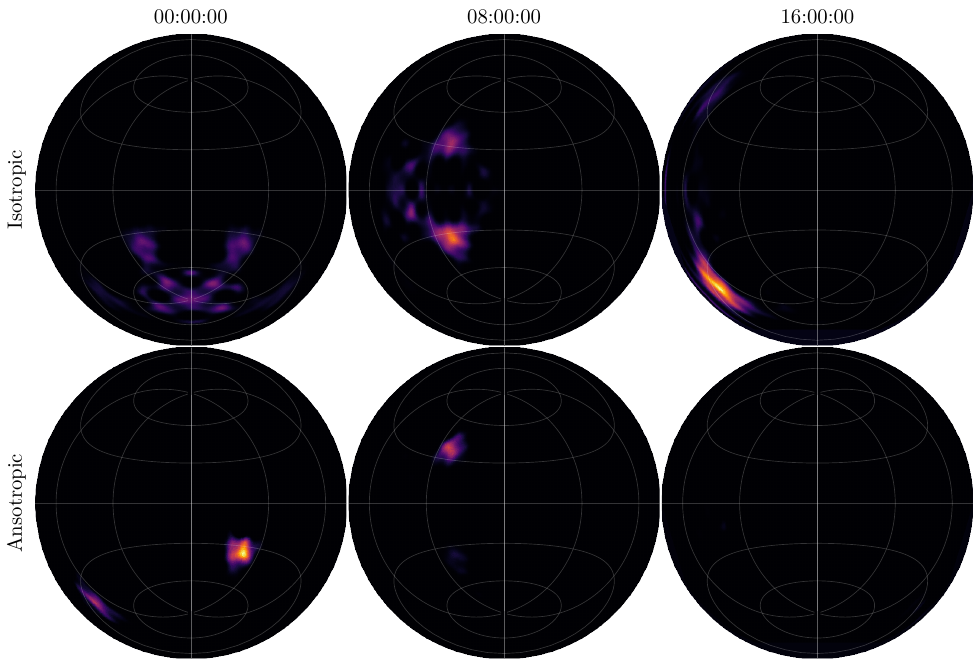}
\caption{Angular distribution of dark matter events of a 340 MeV particle in a germanium crystal at 46.4719\degree N, 81.1868\degree E, at different times for June 1st, 2023, normalized to the day's maximum rate, using a Lambert azimuthal equal-area projection.}
\label{fig:angular}
\end{figure}

The most substantial difference, however, appears when we compare the daily modulation amplitudes in figure \ref{fig:fourier} to the yearly modulation amplitudes in figure \ref{fig:fourieryearly}. In this comparison we see that the daily modulation in the anisotropic component relative to its yearly modulation is significantly higher than in the isotropic component. This can be attributed to the angular distribution of the events being significantly more localized in the anisotropic component than in the isotropic component. This effect is demonstrated in figure \ref{fig:angular}, which shows the angular event rates of the components at different times of day. We show only nontransverse distributions since the effect is quantitatively similar in corresponding transverse distributions.

The analysis here suggests that 
it is in principle possible to infer 
directional characteristics of DM--nucleus scattering from a crystal detector with no directional detection capabilities: knowledge of the orientation of the crystal lattice in the detector and the daily modulation information of the signal allows us to deduce information about the shape of the DM velocity distribution and about the details of DM--nucleus interactions.
One might be interested in understanding to what extent a reconstruction of the directional signal would be possible if modulation data from multiple differently oriented crystals was given, but such an inversion problem is beyond the scope of this work. Instead, in the next section we focus on the simpler problem of understanding quantitatively to which extent this information allows one to distinguish the presence of an anisotropic component in the velocity distribution.

\section{Sensitivity to SHM$^{++}$ with event rate modulation} 
\label{sec:sensitivity}

We wish to find the experimental sensitivity of a germanium ionization detector for the presence of the anisotropic component in the Milky Way halo as a function of the DM mass. For a given DM mass and anisotropic fraction $\eta$ we compute the expected event rate using equation \eqref{truerate} with the distribution \eqref{eq:SHM++}. Because the SHM$^{++}$ distribution is just a linear combination of the isotropic and anisotropic components, we can express the overall event rate as 
\be R=(1-\eta)R_\text{I}+\eta R_\text{A},
\ee 
where $R_\text{I}$ and $R_\text{A}$ are reference event rates for the isotropic and anisotropic components, respectively. From this expected rate we create a binned event distribution using time and energy bins as specified below, by drawing a number of events for each bin from Poisson distribution with the expected number of events given by the expected event rate. We then find the maximum likelihood for this binned event rate to be drawn from the background only ($\eta = 0$) model, and for it to be drawn from the model containing the signal ($\eta \neq 0$). The likelihood is given by
\be
\mathcal{L}(N_\text{I}, N_\text{A}) = e^{
-\sum\limits_{i=1}^N \left( N_\text{I}n_i^\text{I}+N_\text{A}n_i^\text{A} \right)} \prod\limits_{i=1}^N \frac{1}{n_i^\text{obs}!}\left( N_\text{I}n_i^\text{I}+N_\text{A}n_i^\text{A} \right)^{n_i^\text{obs}},
\ee
where $N_\text{I}$ and $N_\text{A}$ are the normalizations of the isotropic and anisotropic component, respectively, 
$N$ is the number of bins, $n_i^\text{I}$ is the expected number of events in bin $i$ from the isotropic component, $n_i^\text{A}$ is the expected number of events in bin $i$ form the anisotropic component and $n_i^\text{obs}$ is the observed number of events in bin $i$ in the simulated experiment, i.e. the number drawn from the model distribution as explained above.

As the time averaged rates $\mean{R_\text{I}}$ and $\mean{R_\text{A}}$ are not equal, the overall time averaged rate $\mean{R}$, and by extension the total number of expected events, depends on the value of $\eta$. Since we want to test sensitivity to the presence of an anisotropic component under the assumption that the DM--nucleus cross section is not known, we normalize the data to some constant number of expected events. That is, we use the expression
\begin{equation}
R = \frac{n_0}{\Delta t}\frac{(1-\eta)R_\text{I}+\eta R_\text{A}}{(1-\eta)\mean{R_\text{I}}+\eta\mean{R_\text{A}}},
\end{equation}
where $n_0$ is the given number of expected events, and $\Delta t$ is the time interval over which the analysis is performed. The expected binned event counts $n^\text{I}_i$ and $n^\text{A}_i$ are then computed for the normalized event rates, such that
\begin{equation}
\sum_{i=1}^N(n^\text{I}_i + n^\text{A}_i)=n_0.
\end{equation}
This ensures we are only testing sensitivity to variations resulting from changes in the temporal modulation and energy spectrum of the events rather than changes in the overall number of events detected.

The test statistic $q_0$ is given by the logarithm of the maximum likelihood ratio
\be
q_0 = 2\log \left( \frac{\max \mathcal{L}(N_\text{A},N_\text{I})}{\max \mathcal{L}(0,N_\text{I})}  \right),
\ee
where the maximum is over varying values of $(N_\text{A},N_\text{I})$, or just $N_\text{I}$ for the $\text{signal}+\text{background}$ and background-only models respectively. For each DM mass and $\eta$ in the signal model we generate 2000 pseudoexperiments by drawing the binned event rates as described above, and find the corresponding value of the test statistic. If $q_0>9$ in more than 90\% of the generated data sets, we conclude that the presence of an anisotropic component is detectable within the confidence limit of three standard deviations. The $3\sigma$ sensitivity of the experiment for a given DM mass is then defined as the limiting value of $\eta$ for which the model remains within discovery reach.

\begin{figure}
\includegraphics{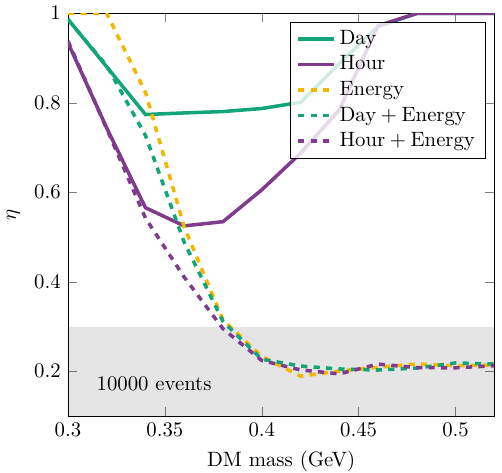}
\includegraphics{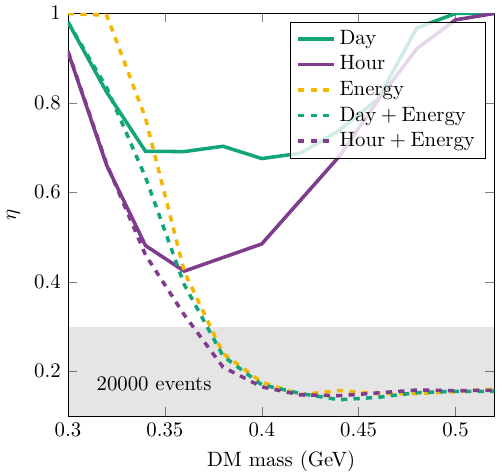}
\caption{Discovery reach of $\eta$ as a function of DM mass in a hypothetical Germanium experiment with 10000 and 20000 signal events. The reach is shown with no time binning (yellow), day long bins (green), and hour long bins (purple), as well as without and with energy binning (solid and dashed lines, respectively) with bin width of 10 eV. The gray region indicates the limits $[0.1,0.3]$ on the magnitude of $\eta$ in the SHM$^{++}$ model.}
\label{fig:reach}
\end{figure}

The resulting reach for various time and energy bins is shown in figure \ref{fig:reach} for the cases $n_0=10000$ and $n_0=20000$. It is notable that because the detector anisotropy here is relevant only for events with energies close to the threshold, its effects likewise are significant only for DM masses close to the lowest mass that can produce events above the threshold energy. In the case of germanium, the relevant mass range is around 300--500 MeV. Indeed, we see that for masses approaching 500 MeV, the time-binning-only reach approaches $\eta=1$ as the contribution of the detector anisotropy on the temporal modulation vanishes. When energy information is included, the discovery reach instead plateaus around $\eta=0.2$ as the isotropic and anisotropic components of the velocity distribution produce distinguishable recoil energy spectra. The impact of the temporal modulation is most significant for DM masses below 400 MeV where the inclusion of daily modulation information improves the reach relative to the energy-binning-only case. For the region $\eta=0.2\pm0.1$ expected by the SHM$^{++}$ model the improvements are smaller and relevant in a narrow mass range.

\begin{figure}
\includegraphics{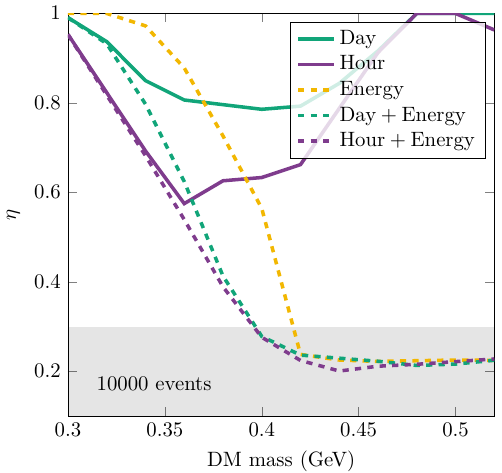}
\includegraphics{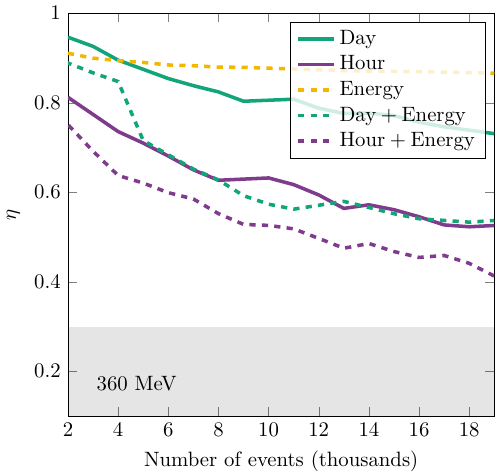}
\caption{Discovery reach of $\eta$ as a function of DM mass in a hypothetical Germanium experiment, marginalized over a choice between $\mean{|\mathcal{M}|^2}\sim 1$ and $\mean{|\mathcal{M}|^2}\sim (\vec{v}_\perp)^2$ DM interaction. Left panel: as a function of mass for 10000 events. Right panel: as a function of event count for mass of 360 MeV.}
\label{fig:reach_mixed}
\end{figure}

The potential to distinguish an SHM$^{++}$-like distribution from a plain symmetric distribution via the temporal modulation of the event rate is a result of the anisotropic component having a different angular distribution of events. However, the transverse and nontransverse Radon transforms in equation \eqref{eq:radon} likewise have different angular distributions. This is because the presence of the $(\vec{v}_\perp)^2$ term in the transverse case decreases the number of events with recoil directions parallel to the DM velocity, which results in a ring-like shape for the distribution of scattering events, instead of a peaked shape as we have in the nontransverse case. It is therefore reasonable to ask whether our ability to detect the presence of an anisotropic component in the velocity distribution is weakened by ignorance of the underlying DM model. To that end, we performed a likelihood analysis similar to the one described above, but this time the likelihood ratios were marginalized over a choice between a transverse and nontransverse DM interaction. That is, we assumed there to be a 50\% chance for either $\mean{|\mathcal{M}|^2}\sim 1$ or $\mean{|\mathcal{M}|^2}\sim (\vec{v}_\perp)^2$. The resulting reach for 10000 events is shown in the left panel of figure \ref{fig:reach_mixed}. Comparing to figure \ref{fig:reach}, we see that the reach is weakened for masses below 420 MeV, especially in the case with no time binning, as well as for higher masses in the cases where no energy information is given. This is consistent with the fact that the temporal modulation from an anisotropic component is somewhat hard to distinguish from a modulation due to a non-identity DM interaction, whereas the energy spectra of these different scenarios are distinct. The right panel of the figure shows how the sensitivity scales with the number of events for a selected dark matter particle mass of 360 MeV. We observe that the relative enhancement of the sensitivity due to the daily modulation remains quite stable, but on absolute scale the sensitivity weakens so that the observationally motivated range $\eta\lesssim0.3$ is not within reach for lower number of events. However, this behaviour indicates that the daily modulation signal could remain useful for identifying different structures in the velocity distribution even at lower event numbers, in cases where the anisotropic component is expected to be larger.

\section{Conclusions}
\label{sec:checkout}

A simple benchmark case for the DM velocity distribution in galaxies 
is given by the SHM: a Maxwellian distribution with a cutoff at the local escape velocity.  
However, it is unlikely that this would generally hold.
For the Milky Way, the observations by the GAIA satellite~\cite{Gaia:2018ydn} on the motion of stars, suggest an anisotropic 'sausage' component~\cite{Evans:2018bqy}, leading to the introduction of an SHM$^{++}$ model with a highly anisotropic component to the dark matter velocity distribution. The effect of this additional component is to increase the portion of dark matter with large radial and small tangential velocities in the galactic frame. 

In this paper we have shown that, apart from having a distinct recoil energy spectrum, for high-enough values of the anisotropy parameter $\beta$ the anisotropic component leads to a unique bimodal angular distribution of recoil events. In an ionization detector with an anisotropic response this leads to a daily modulation signal that is distinct from the one coming from an isotropic velocity distribution. We carried our analysis out concretely for a Germanium detector and determined several periodic patterns arising from the crystalline structure of the target. This analysis was performed under the assumption that the directional dependence of the ionization threshold matches that of the defect creation threshold, which is calculable using classical molecular dynamics simulations. While this assumption might not exactly hold in reality, it is likely that the behaviour is similar as both effects originate from the same crystal structure of the material. In the absence of an alternative model for the ionization threshold, it is difficult to quantitatively evaluate the level of uncertainty in our results related to this choice. In the future we hope that improved simulations and experimental data will be able to confirm the validity of our approach.

We then carried out a likelihood ratio test analysis to determine the experimental sensitivity to the anisotropy fraction $\eta$. We found that with a sufficiently large number ($\mathcal{O}(10^4)$) of signal events, and for DM masses for which most of the recoil spectrum is above the detector threshold, an experiment with $<10$ eV energy resolution can detect the presence of an anisotropic component with an anisotropy fraction allowed in the SHM$^{++}$ model. For DM masses that fall in the narrow range near the detection threshold substantial gains in sensitivity to the anisotropic component can be made with daily modulation information, but this isn't sufficient to probe the SHM$^{++}$ region.

It can be expected that qualitatively similar results arise for other similar target materials like Silicon~\cite{Dinmohammadi:2023amy}, and for different anisotropic velocity distributions. It would be interesting to extend this work to other materials for which directional sensitivity is expected~\cite{Sassi:2022njl}.

\acknowledgements{The financial support from the Research Council of Finland (grant\# 342777) is gratefully ackowledged. S.S. acknowledges financial support from Magnus Ehrnrooth foundation.}

\bibliography{bibliography.bib}

\end{document}